\documentclass[twocolumn,showpacs,prl]{revtex4}

\usepackage{epsfig}
\usepackage{amsmath}
\usepackage{graphicx}
\usepackage{dcolumn}
\usepackage{amsmath}
\usepackage{latexsym}

\begin{document}

\title{The effect of a weak ferromagnetic matrix on a system of nanomagnetic particles}

\author{S. Chakraverty$^{1}$, A. Frydman$^{2}$, V.G. Pol$^{3}$, S. V. Pol$^{3}$  and A. Gedanken$^{3}$ \\
} \affiliation{ 1. Nanoscience Unit, S. N. Bose National Center
for Basic Sciences, Block-JD, Sector-III, Salt Lake,
Kolkata - 700098, India.\\
2. Dept.of Physics, Bar Ilan University, Ramat Gan 52900, Israel
\\
3. Department of Chemistry and Kanbar Laboratory for Nanomaterials
at the Bar-Ilan University Center for Advanced Materials and
Nanotechnology, Bar-Ilan University, Ramat-Gan, 52900, Israel.}

\begin{abstract}
The study of system of magnetic nano-particle has received
increasing attention recently both because of the novel physical
concepts involved and also because of their vast potential for
application. The influence of background material (the substrate
coating) on magnetic properties of such systems is a relatively
open topic and often a full understanding is missing. In the
present work we discuss our experiments and interpretation for two
systems: Ni nanoparticles coated with graphitic carbon and Ni
nanoparticles coated with Au. While the latter system exibits
behavior typical of superparamagnetic particle systems the former
shows several puzzling results such as extremely high blocking
temperature ($T_{B}$), very fast relaxation time well below
$T_{B}$, temperature independent field-cooled magnetization and
very small coercivity and remanent magnetization. We interpret
these findings as being a result of weak ferromagnetism,
characteristic of the graphitic carbon. This induces strong
magnetic interactions between the Ni particles in the presence of
small magnetic fields. Such systems give rise to a dramatic
difference in blocking temperature between measurements performed
at zero field and those performed at very small magnetic fields.

\end{abstract}
\date{\today}

\maketitle

Systems of magnetic nanoparticles have been gaining increasing
attention over the past few years both because they introduce
novel physical concepts and because of their vast potential for
applications in the fields of nano-electronics, storage media and
medicine \cite{nano, nano0, nano1, nano2, nano3, nano3a,nano4}.
For practical devices, the nanoparticles are usually imbedded in a
matrix or placed on a solid substrate. The influence of different
background materials on the magnetic properties of the
nanoparticle system is a relatively open topic. Though it is clear
that different materials can have various effects on the magnetic
coupling between the particles (depending on the magnetic nature
of the matrix) a full understanding of these issues is lacking.

In this letter we study the effect of a special matrix material,
i.e. a weak ferromagnet. We compare the magnetic behavior of Ni
nanoparticles embedded in gold (a diamagnetic material) to that of
Ni particles embedded in graphitic carbon which has been shown to
exhibit weak ferromagnetic properties \cite {carbon1,carbon2}. The
latter exhibit a set of surprising and contradictory findings.
These are attributed to the coupling between the nanomagnetic
particles via the weak ferromagnetic material in the presence of a
small magnetic field. We show that this is an unique case in which
the magnetization and blocking temperature of the system strongly
depend on whether a small external field is applied.

The synthesis of Ni-C (graphitic) core-shell nanostructures was
carried out by thermal dissociation of nickel acetylacetonate,
Ni(C5H7O2)2, in a closed vessel cell assembled from stainless
steel Swagelok parts. The full process is described elsewhere
\cite{vilas}. Due to the lack of a commercially available gold
precursor to carry out a similar synthesis, we produced Ni-Au
core-shell using a two step process. In the first, the nuclei of
Ni nanoparticles were formed in a solution of nickel (II) acetate
tetrahydrate [99.998, Aldrich Chemical Co.] in  EG (Ethylene
glycol). A 100mL glass flask was placed in a microwave oven
(spectra 900W) that was connected to the water condenser. A 50 mL
of a solution of $0.2M Ni(Ac)_{2}$ in EG was purged by argon for
10 min, after which the microwave oven was turned on at the power
level of 60 watt with a continued flow of the argon gas. A black
suspension appeared after 15 minutes of microwave irradiation. In
the second stage, 0.15mL chloroauric acid (HAuCl4) was added by a
microliter size syringe (Hamilton Co.) in the black suspension of
Nickel nanoparticles. Once again microwave irradiation was turned
on and maintained for 20 minutes at a power of 60 watt. The
resulting pale-black solid product was washed thoroughly with
ethanol thrice and centrifuged at 8000 rpm. The as-prepared
product was dried in a vacuum chamber for 12h.

The morphology and crystal structure for both samples were studied
by TEM and HR-TEM. Figure 1a depicts the TEM analysis of a Ni
nanoparticle (with sizes ranging from 30-150nm) embedded in
graphitic carbon. The inserted top right image confirms the
formation of ordered graphitic carbon shell of ~15nm. The
interlayer spacing between these graphitic planes is 3.41 \AA,
which is very close to that of the graphitic layers. The inserted
bottom image provides further verification for the identification
of the core as Ni. It illustrates the perfect arrangement of the
atomic layers and the lack of defects. The measured distance
between these (111) lattice planes is 0.200 nm, which is very
close to the distance between the planes reported in the
literature (0.203 nm) for the face-centered cubic lattice of the
Ni ((PDF No. 03-065-2865).

The TEM analysis of Ni particles (sizes of 100-200nm) coated by
gold nanocrystals with diameters of 7nm is shown in Figure 1b. The
inserted (HR-TEM) top image confirms the interlayer spacing for Au
nanocrystals and the bottom image provides further evidence for
the Ni core.

\begin{figure}\vspace{-0.8cm}
\centerline{\epsfxsize=2.8in \epsffile{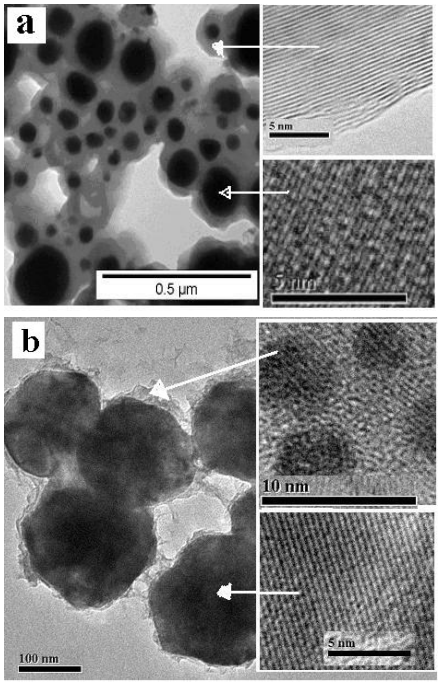}}
\vspace{-0.5cm} \caption{Transmission electron micrographs of (a)
Ni nanoparticles embedded in graphitic carbon, and (b) Ni
particles coated by gold nanocrystals. The left panels are
conventional low magnification TEM images and the right panels are
high resolution TEM images taken at different sections of the
samples. } \vspace{-0.5cm}
\end{figure}

Fig. 2 depicts the field cooled - zero field cooled (FC -ZFC)
magnetization measurements for the Ni-C and Ni-Au samples as
measured in a SQUID magnetometer. In ZFC mode the system is cooled
to the lowest temperature in the absence of magnetic field, H, and
M(T) is measured while heating the sample in the presence of
H=100Oe, while in FC, M(T) is measured during cooling the sample
in the presence H=100Oe. For superparamagnetic systems the ZFC
magnetization curves are expected to show a peak corresponding to
the average blocking temperature while the FC curves always
increase as the temperature is decreased due to the alignment of
the spins in the direction of the field. Our measurements show
that for the Ni-Au system $T_{B}$ is found to of the order of 150K
(which is a reasonable value for Ni particles with diameters of
~150 nm). Since the Ni particles in the carbon coated systems are
smaller, the average blocking temperature is expected to be
smaller as well (a rough estimation yields $\sim 30K$).
Surprisingly, for the Ni-C case the blocking temperature is found
to be well above 400K (which was the highest available
temperature). Moreover, the temperature dependance of the
magnetization in these samples is very peculiar. The FC C-Coated
Ni particles exhibit magnetization that remains almost constant
even down to temperatures of 15K.

\begin{figure}\vspace{-0.8cm}
\centerline{\epsfxsize=2.8in \epsffile{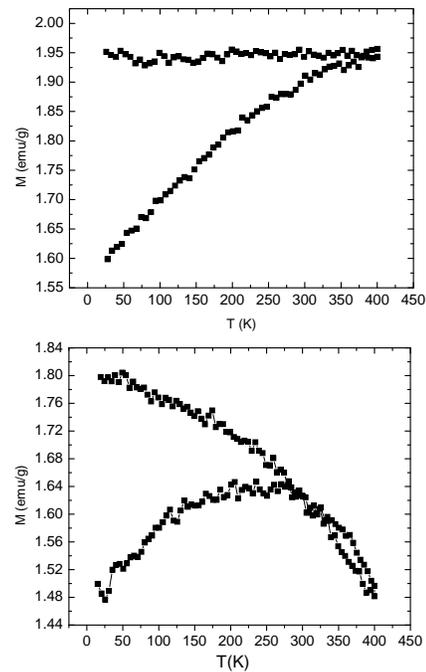}} \vspace{-0.8cm}
\caption{FC and ZFC M(T) plots for Ni-C (top) and Ni-Au (bottom)
nanoparticle systems.} \vspace{-0.5cm}
\end{figure}

Another peculiarity of these systems is demonstrated in fig. 3
which depicts the low-field regime of the room temperature M-H
curves of both samples. It is seen that the coercive field,
$H_{C}$ is \emph{smaller} for the carbon coated sample $(\sim 60
Oe)$ than that of the gold coated sample $(\sim 170 Oe)$. This
seems to be in striking contradiction to the fact that the
blocking temperature extracted from the FC-ZFC measurements is
much larger for the Ni-C sample.
\begin{figure}\vspace{-0.8cm}
\centerline{\epsfxsize=2.8in \epsffile{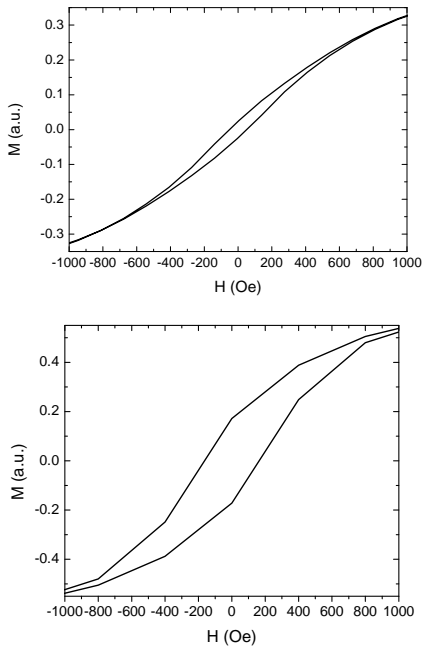}} \vspace{-0.8cm}
\caption{Room temperature hysteresis loops for the Ni-C (top) and
the Ni-Au (bottom) samples} \vspace{-0.5cm}
\end{figure}

An even more surprising finding is related to relaxation-response
measurements below the blocking temperature. In this measurement,
the sample is cooled to a temperature of 40K (well below the
measured $T_{B}$ for both types of samples) in the absence of
magnetic field. At this temperature a 50 Oe field is abruptly
switched on and the magnetization is measured as a function of
time. After a few hours the field is switched back off and the
M(t) is measured again. Figure 4 compares the relaxation-response
measurements for the Ni-Au and Ni-C samples. It is seen that the
Au coated Ni sample shows slow response to changes in the external
magnetic field. When the field is switched on the magnetization
increases slowly and and when it is switched back off there is a
long time related to the magnetization decay. This is expected
because the magnetic orientation of a particle is blocked at
temperatures smaller than $T_{B}$  . The Ni-C sample, on the other
hand, responds almost immediately (within the measuring time scale
of SQUID) to the magnetic field changes. Upon switching the field
off the magnetization drops immediately to zero and upon switching
the magnetic field back on the magnetization restores its maximal
value instantaneously. This behavior is very unexpected for a
system of particles with blocking temperatures above 400K.

In an attempt to understand the results observed in the Ni-C
samples we considered the possibility that the large obtained
blocking temperatures are due to dipole-dipole interactions
between the magnetic moments of the Ni-nanoparticles
(super-spin-glass behavior \cite{petra}) mediated by the
ferromagnetic background. To check the relevance of this mechanism
in our system we performed arrested cooling ZFC measurements. In
this experiment the system is cooled in the absence of magnetic
field from room temperature to 30K (well below the $T_{B}$ of the
system) and the temperature is arrested for four hours. Then the
system is cooled again in a constant rate to 15K in absence of
external magnetic field. From 15K the sample is heated up to room
temperature in the presence of 50Oe DC external magnetic field,
while monitoring the magnetization constantly. An established sign
of super-spin-glass interactions \cite{petra} is the occurrence of
a dip in the M-T heating curve at the temperature that was
arrested during the cooling process (30K in our case). Our Ni-C
samples showed no sign for such a dip, thus ruling out the
existence of pronounced spin glass interactions.

\begin{figure}\vspace{-1cm}
\centerline{\epsfxsize=2.98in \epsffile{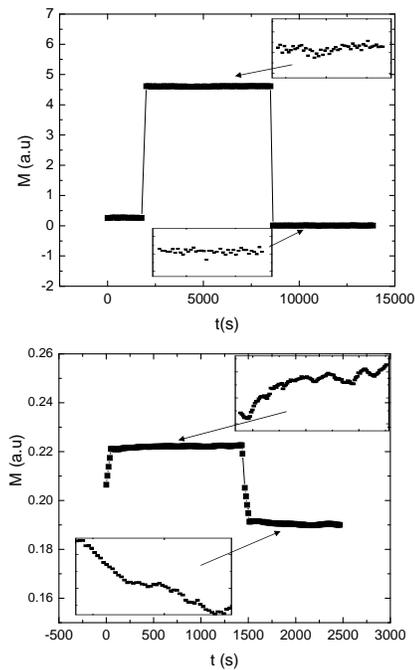}}
\vspace{-0.8cm} \caption{relaxation response graphs for Ni-C (top)
and Ni-Au (bottom) nanoparticle systems.} \vspace{-0.5cm}
\end{figure}

Nevertheless, the fact that the puzzling (and apparently
contradicting) results were not observed in the Ni-Au systems
leads us to suggest that the unique phenomena are due to
interparticle (Ni-Ni) interaction mediated via the weak
ferromagnetic carbon matrix. In the following paragraphs we
suggest a model, based on such interactions, that may account for
our experimental findings.

Let us consider an assembly of Ni nanoparticles coated by
graphitic carbon, in which the easy axis of the Ni-particles are
randomly oriented. The carbon is characterized by very weak
ferromagnetism. Let $E_{nn}$ be the magnetic interaction between
the atoms of Ni, $E_{nc}$ the interaction between Ni and C and $
E_{cc}$ the interaction energy between carbon atoms. The dipolar
interaction between Ni-nanoparticles is assumed to be weak due to
small Ni particle size and large inter-particle separation. Let us
further assume that $E_{nc }>> E_{cc}$. In the absence of magnetic
field there is a spin disorder in the carbon shell coating the Ni
particle due to the surface spin disorder of the Ni particles
\cite{kodoma} and the random orientation of the easy axis. But as
soon as an external magnetic field is applied, all the Carbon
spins orient themselves in the field direction, producing a huge
effective magnetic field on the Ni-nanoparticle, H', thus forcing
it to orient along the external field direction.

The apparent contradiction observed for Ni-C samples occurs due to
the fact that two different classes of experiments (namely field
measurements and zero field measurements) were compared. FC-ZFC
belongs to finite applied field measurements while relaxation
measurements belong to zero field measurements. Taking this fact
into account, our experimental observations can be well explained
within the framework of the above model. Let us first discuss the
FC-ZFC measurement. Consider a single domain uniaxial magnetic
nanoparticle. The anisotropy energy of such a particle is given
by:
\begin{equation}
E = KV sin^{2}\theta
\end{equation}
where K is the anisotropy energy per unit volume, V is the volume
of the nanoparticle and $\theta$ is the angle between the magnetic
moment of the particle and the easy axis for magnetization. In
absence of an external magnetic field the probability of finding
an up spin is equal to that of a down spin, leading to zero
magnetization. If the temperature of the system is decreased  in
absence of the external magnetic field the situation remains the
same and zero magnetization is observed. As soon as an external
magnetic field is applied the energy of the system becomes
\begin{equation}
E = KV sin^{2}\theta-\mu VH^{\star} cos\theta
\end{equation}

where $\mu$ is the magnetic moment of the particle per unit volume
and $H^{\star}=H+H'$ is the effective magnetic field seen by the
particle due to external magnetic field and the field produced by
surrounding Carbon spins. In eq. 2 we have assumed, for
simplicity, that the external magnetic field is applied along the
direction of easy axis of the nanoparticle. Let $\delta E_{0m}$ be
the energy difference between the minima at $\theta=0$ and that of
the maxima $E_{m}$, and $\delta E_{1m}$ be that between the energy
at $\theta=\pi$ and $E_{m}$. Due to the contribution of large
number of Carbon spins, $H^{\star}$, is extremely large. This
leads to $\delta E_{0m} \gg \delta E_{1m}$, which means that with
increasing temperature the down spin will easily orient parallel
to the applied magnetic field with the help of thermal energy. The
magnetization can decrease with increasing temperature only if the
thermal energy becomes comparable to $\delta E_{0m}$, which is a
very high value. Hence, unusually large effective blocking
temperature are measured by the ZFC magnetization experiment. If
now the system is cooled in the presence of an external magnetic
field, the particles are already blocked along the direction of
external field and the FC curve becomes independent of
temperature.

The magnetization response measurement are also understood using
the above considerations. As soon as the magnetic field is switch
on the system's hamiltonian becomes that of eq. 2. Since
$H^{\star}$ produces a large asymmetry in the energy profile. In
this case the response time (the time it takes for the spins to
flip along the external field)given by:

\begin{equation}
\tau=\tau_{0}exp[\frac{KVsin^{2}(\theta)+\mu
VH^{\star}cos(\theta)}{K_{B}T}]
\end{equation}
is very short compared to that of the Ni-Au samples in which the
magnetic field is much smaller.

The magnetization relaxation measurement, on the other hand, is a
zero field magnetic measurements. In effect these systems are
characterized by two distinct blocking temperatures; a zero field,
conventional blocking temperature, $T_{B}(0)$, and a finite field
blocking temperature $T_{B}(H)$, which is much larger due to the
effect of the carbon atoms. When the magnetic field is switched
off, the carbon spins disorient immediately. Since $T_{B}(0)$ of
the Ni particles is estimated to be a few tens of degrees, at 40K
they behave as isolated nano-magnets with a small barrier height.
Hence, fast relaxation is expected. This is in vast contradiction
to the situation for the FC-ZFC measurements in which a field is
applied and the finite field blocking temperature, $T_{B}(H)$,
determines the magnetization properties. A similar argument is
related to the coercive field, $H_{C}$ which is the field
associated with zero magnetization. Since very small external
magnetic fields are sufficient to align the grains' orientations,
thus giving rise to large measured magnetization, the field for
which zero magnetization is achieved in the Ni-C samples is
expected to be very low.

In conclusion we note the the choice of substrate and spacing
material is crucial for the magnetic performance of nano-particle
based systems. A graphitic carbon spacer gives rise to very fast
response to magnetic field changes while exhibiting relative
insensitivity to temperature changes.

We thank R. Ranganathan for useful discussions. This research was
supported by the Israel Science foundation (grant number 326/02).

\end{document}